\documentclass[aps,prl,showpacs,10pt, superscriptaddress,twocolumn,byrevtex,floatfix]{revtex4-1}
\usepackage{amsmath}
\usepackage{graphicx}
\usepackage{amsmath}
\usepackage{braket}
\usepackage{color}
\usepackage{hyperref}
\usepackage{tikz}
\usepackage{array}

\hypersetup{bookmarksnumbered,%
	colorlinks,%
	linkcolor=blue,%
	citecolor=blue,%
	plainpages=false,%
	pdfstartview=FitH}

\newcommand{\nc}{\newcommand}           
\nc{\vc}[1]     {\mbox{\boldmath $#1$}} 
\nc{\mapleft}[1]{                       
 \smash{\mathop{                        %
  \hbox to 0.90cm{\rightarrowfill} }\limits_{#1}}}
\nc{\figwidth}{0.55}                    

\nc{\mydraft}	{\setlength{\topmargin}{-1.5cm}}
\mydraft

\begin{document}

\title{Cluster-breaking and reconfiguration effects in $_\Lambda^{12}\rm{B}$ hypernucleus}

\author{Jiaqi Tian}
\affiliation{College of Physics, Nanjing University of Aeronautics and Astronautics, Nanjing 210016, China}
\affiliation{KEK Theory Center, Institute of Particle and Nuclear Studies (IPNS), High Energy Accelerator Research Organization (KEK), 1-1 Oho, Tsukuba, Ibaraki, 305-0801, Japan}
\affiliation{Key Laboratory of Aerospace Information Materials and Physics (NUAA), MIIT, Nanjing 211106, China}

\author{Mengjiao Lyu}
\email{mengjiao.lyu@nuaa.edu.cn}
\affiliation{College of Physics, Nanjing University of Aeronautics and Astronautics, Nanjing 210016, China}
\affiliation{Key Laboratory of Aerospace Information Materials and Physics (NUAA), MIIT, Nanjing 211106, China}

\author{Akinobu Doté}
\affiliation{KEK Theory Center, Institute of Particle and Nuclear Studies (IPNS), High Energy Accelerator Research Organization (KEK), 1-1 Oho, Tsukuba, Ibaraki, 305-0801, Japan}
\affiliation{J-PARC Branch, KEK Theory Center, IPNS, KEK, 203-1, Shirakata, Tokai, Ibaraki, 319-1106, Japan}
\affiliation{Graduate Institute for Advanced Studies, SOKENDAI, 1-1 Oho, Tsukuba, Ibaraki, 305-0801, Japan}

\author{Zheng Cheng}
\affiliation{College of Physics, Nanjing University of Aeronautics and Astronautics, Nanjing 210016, China}
\affiliation{Key Laboratory of Aerospace Information Materials and Physics (NUAA), MIIT, Nanjing 211106, China}

\author{Takayuki Myo}
\affiliation{Research Center for Nuclear Physics (RCNP), Osaka University, Osaka 567-0047, Japan}
\affiliation{General Education, Faculty of Engineering, Osaka Institute of Technology, Osaka, Osaka 535-8585, Japan}

\author{Masahiro Isaka}
\affiliation{Hosei University, 2-17-1 Fujimi, Chiyoda-ku, Tokyo 102-8160, Japan}

\author{Hisashi Horiuchi}
\affiliation{Research Center for Nuclear Physics (RCNP), Osaka University, Osaka 567-0047, Japan}

\author{Hiroki Takemoto}
\affiliation{Faculty of Pharmacy, Osaka Medical and Pharmaceutical University, Takatsuki, Osaka 569-1094, Japan}

\author{Hiroshi Toki}
\affiliation{Research Center for Nuclear Physics (RCNP), Osaka University, Osaka 567-0047, Japan}

\author{Niu Wan}
\affiliation{School of Physics and Optoelectronics, South China University of Technology, Guangzhou 510641, China}

\author{Qing Zhao}
\affiliation{School of Science, Huzhou University, Huzhou 313000, Zhejiang, China}

\date{\today}


\begin{abstract}
  We investigate the cluster-breaking effect and spatial distribution of negative-parity states in the $_\Lambda^{12}\rm{B}$ hypernucleus using the Hyper-Brink model with cluster-breaking (CB-Hyper-Brink) optimized via Control Neural Network (Ctrl.NN). The results demonstrate that the inclusion of cluster-breaking is essential for accurately reproducing the observed low-lying energy levels and for making reliable predictions of the Hoyle-analog state $1_4^-$ in $^{12}_\Lambda$B. Cluster-breaking manifests as strong spin-orbit correlations and the dissolution of ideal cluster configurations, as revealed by the analysis of one-body spin-orbit operator expectation values and the spatial overlap with projected cluster bases. 
  The interplay between short-range repulsion and 
  intermediate-range attraction in the $\Lambda N$ interaction induces the cluster reconfiguration effect, which is characterized by the coexistence of $\Lambda$-$\alpha$ and $\Lambda$-triton correlations; this reconfiguration effect leads to a modest stabilization and shrinkage of cluster structures. The variation in electric quadrupole transition strengths, $B(E2)$, between the ground and Hoyle-analog states serves as a sensitive probe for the degree of cluster-breaking, providing direct evidence for its physical relevance. These findings highlight the crucial role of cluster-breaking in characterizing the hypernuclear structure and offer a comprehensive framework for understanding the interplay between clustering and shell-model dynamics in hypernuclei.
\end{abstract}
\maketitle


\section{I. Introduction}

The atomic nucleus represents a complex fermionic many-body system composed of protons and neutrons interacting via intricate nuclear forces \cite{Ebran2012, KanadaEnyo2012}. Depending on the excitation energy and nuclear composition, nuclei can exhibit structures ranging from strongly correlated cluster states to independent-particle shell-model-like configurations \cite{Ye2024,KanadaEnyo2012}. Typically, low-lying nuclear states are successfully described using shell models with prominent mean-field effects, while states near particle emission thresholds frequently manifest cluster structures, where nucleons form well-defined subunits, notably $\alpha$ clusters \cite{freer_microscopic_2018}. The delicate balance among medium-range nuclear attraction, antisymmetrization (Pauli blocking), and short-range repulsion is fundamental to the formation and dissolution of these clusters, which are an interplay often termed as cluster-shell competition \cite{freer_microscopic_2018}.

In particular, the $\alpha$ particle ($^4$He), owing to its remarkable stability and strong binding, serves as a crucial building block in cluster models \cite{freer_microscopic_2018}. Numerous theoretical frameworks, including traditional Brink-type cluster models, antisymmetrized molecular dynamics (AMD), and the antisymmetrized quasi-cluster model (AQCM), have been developed to describe such cluster structures \cite{zhou2013,Funaki2015,zhou_2_2018,Lyu2016,cheng_evidence_2024,KanadaEnyo2012, Kimura2016, Ono2004, suhara_cluster_2012,itagaki_simplified_2005,masui_simplified_2007,yoshida_appearance_2009,itagaki_simplified_2011,itagaki_simplified_2012, Itagaki2018,Itagaki2022,itagaki_cluster-shell_2023,myo_variation_2023, Myo2024}. The AMD approach, which treats nucleons individually without assuming clusters a priori, naturally captures cluster and shell-model features within a unified framework \cite{KanadaEnyo2012, Kimura2016, Ono2004}. AQCM explicitly bridges cluster and shell structures by introducing an adjustable parameter controlling cluster dissolution, effectively simulating the transition from well-defined cluster states to single-particle configurations \cite{itagaki_simplified_2005,masui_simplified_2007,yoshida_appearance_2009,itagaki_simplified_2011,itagaki_simplified_2012,Suhara2013, Itagaki2018,Itagaki2022,itagaki_cluster-shell_2023}. These models have successfully described phenomena such as the Hoyle state in $^{12}$C—a dilute 3$\alpha$ configuration—and provided valuable insights into how nuclear forces, particularly spin-orbit interaction, drive cluster-breaking and promote shell-model structures.

The spin-orbit interaction plays a crucial role in determining the balance between cluster and shell structures in nuclei. Physically, the spin-orbit force favors states where nucleons occupy distinct single-particle orbits aligned with their spin orientation, thereby disrupting the spatial correlations essential for clustering. Theoretical analyses have demonstrated that the spin-orbit force significantly dissolves cluster configurations, resulting in cluster-shell coexistence \cite{itagaki_simplified_2005,masui_simplified_2007,yoshida_appearance_2009,itagaki_simplified_2011,itagaki_simplified_2012, Suhara2013,Itagaki2018,Itagaki2022,itagaki_cluster-shell_2023}. Consequently, elucidating the interplay of these interactions within nuclear structures becomes essential for comprehensively understanding nuclear dynamics and structural evolution across nuclear charts.

This complexity is further enriched in hypernuclei, nuclear systems containing hyperons such as the $\Lambda$ particle. Due to the attractive $\Lambda N$ interaction and the absence of Pauli exclusion between the $\Lambda$ and nucleons, hyperons typically reside in deeply bound $s$-states, inducing a pronounced shrinkage of nuclear matter and potentially altering the balance between cluster and shell configurations \cite{motoba_progr_1983,hiyama_gamma_1999, tanida_measurement_2001,Akikawa2002,hiyama_structure_2009}. Hypernuclei thus provide unique laboratories to investigate nuclear structure dynamics under modified binding conditions. For example, theoretical and experimental studies have demonstrated shrinkage effects in hypernuclei such as $^7_\Lambda$Li and $^{9}_\Lambda$Be, where the presence of the $\Lambda$ particle significantly reduces inter-cluster distances, influencing observables like radii and electromagnetic transition strengths \cite{motoba_progr_1983,hiyama_gamma_1999,tanida_measurement_2001,Akikawa2002,isaka_low-lying_2020,tian_lambda_2024}.

Among hypernuclei, $^{12}_\Lambda$B is particularly compelling as a benchmark system due to its complex nuclear structure, which integrates features of both clustering and shell-model states \cite{yamada__2010,suhara_cluster_2012,zhou_2_2018,Itagaki2022}. Recent theoretical work using advanced computational methods, including neural network-optimized cluster model, has indicated that $^{12}_\Lambda$B may exhibit a rich variety of structural motifs ranging from compact shell-model-like states to extended cluster configurations such as $\alpha + \alpha + t$ chains \cite{Tian2025}. Furthermore, experimental efforts, notably through high-resolution spectroscopy and $\gamma$-ray spectroscopy, have begun to provide precise constraints on the excitation spectra and transition rates of $^{12}_\Lambda$B, facilitating rigorous tests of theoretical predictions \cite{Miyoshi2003,iodice_high_2007,tang_experiments_2014}.

A previous study on $^{13}_\Lambda$C has already highlighted how the presence of a $\Lambda$ particle significantly influences the cluster-shell competition, demonstrating that the $\Lambda$ particle effectively enhances the shell-model characteristics by the cluster dissolution due to spin-orbit interaction \cite{itagaki_cluster-shell_2023}. Building upon these findings, our present work studies the negative-parity states of the analogous and experimentally accessible hypernucleus $^{12}_\Lambda$B, emphasizing the Hoyle-analog-like state and its structural evolution by applying the Hyper-Brink model with cluster-breaking (CB-Hyper-Brink) enhanced by cutting-edge Control Neural Network (Ctrl.NN) optimization \cite{cheng_evidence_2024,tian_lambda_2024,Tian2025} . By closely comparing our predictions with high-precision experimental data, we further aim to clarify the intricate interplay between cluster structure and shell-model states in hypernuclear systems, thereby contributing essential insights into nuclear many-body dynamics under the influence of strangeness.


The paper is organized as follows. In Sec. II, the theoretical framework of the Hyper-Brink model with cluster-breaking (CB-Hyper-Brink) enhanced by Control Neural Network (Ctrl.NN) optimization for hypernuclei is explained. 
In Sec. III, we present the numerical results and discuss the structural properties of $^{12}_\Lambda \rm{B}$ hypernuclei. 
Finally, in Sec. IV, we provide our conclusion.

\section{II. Theoretical Framework}\label{sec:frame}
In this section, we introduce the theoretical framework of Hyper-Brink model with cluster-breaking(CB-Hyper-Brink).
\subsection{A. Hamiltonian}
The Hamiltonian used in the present calculation is composed of the kinetic energy of $\Lambda$ hyperon ($T_\Lambda$) and nucleon ($T_N$), 
nucleon-nucleon central interaction $V_{i j}^{(N N)}$, $\Lambda N$ central interaction $V_i^{(\Lambda N)}$, 
nucleon-nucleon spin-orbit coupling interaction $V_{i j}^{(N N l s)}$ and Coulomb interaction $V_{i j}^{(C)}$,
\begin{equation}
    \begin{aligned}
    H = &\sum_{i=1}^{11} T_i^N+T^{\Lambda}-T_G+\sum_{i<j}^{11} V_{i j}^{(C)}\\
    & +\sum_{i<j}^{11} V_{i j}^{(N N)}+\sum_{i<j}^{11} V_{i j}^{(N N l s)}+\sum_{i=1}^{11} V_i^{(\Lambda N)}.
    \end{aligned}
\end{equation}
We eliminate the effects of spurious center-of-mass motion, denoted as $T_G$, in the Hamiltonian. 
In this paper, we focus on hypernuclear states in which the core nucleus involves the $sd$-shell excitation as well as the ground state and the $\Lambda$ hyperon occupies the $s$-wave state.
Hence, the $\Lambda N$ spin-orbit coupling interaction is so weak that we neglect it here. 
We adopt Volkov No.2  in nucleon-nucleon central force $V_{i j}^{(N N)}$ and set the Majorana parameter $M = 0.60$. 
The G3RS interaction is adopted in the nucleon-nucleon spin-orbit coupling interaction $V_{i j}^{(N N l s)}$. The strength of G3RS interaction is adopted as 1600 MeV.
To compare with cluster model calculation, we also adopt another parameter set.
The Majorana parameter is tuned as $M = 0.59$, and the spin-orbit strength is 2800 MeV.
This parameter set is the same as that used in Ref. \cite{zhou_2_2018}.
In the following results and discussion, this parameter set of interaction is called ``Cluster interaction"

For $\Lambda N$ force $V_i^{(\Lambda N)}$, we adopt ESC14 interaction with many body effects (MBE), which is an effective local interaction with the Gaussian form and simulates the G-matrix calculation with the Nijmegen ESC potential, where the $\Lambda N-\Sigma N$ coupling is renormalized by the G-matrix calculation. In this paper, we adopt the tuned version proposed in Ref.~\cite{isaka_low-lying_2020}, where the spin-dependent part is tuned to reproduce the splitting between $1/2^+$ and $3/2^+$ states of $^{7}_\Lambda\rm{Li}$.

As shown in Ref.~\cite{isaka_low-lying_2020}, the $\Lambda$ binding energies of various 
$\Lambda$ hypernuclei can be systematically reproduced by employing the 
HyperAMD calculation with ESC14+MBE interaction.
In present calculation, we adopt the Fermi momentum parameter $k_F=1.06$ fm$^{-1}$, which is consistent with our previous work in $_\Lambda^{12}\rm{B}$ \cite{Tian2025}.

\subsection{B. Hyper-Brink wave function}
We start with the Hyper-Brink cluster wave function. 
The single particle wave functions are described by Gaussian form functions.
\begin{equation}
  \begin{aligned}
  \phi^{\Lambda, N}(\boldsymbol{r},\boldsymbol{R})&=
      \left(\frac{2\nu_{\Lambda, N}}{\pi}\right)^{3/4}
      \exp\left\{
          -\nu_{\Lambda, N}(\boldsymbol{r}-\boldsymbol{R})^{2}
      \right\}\chi_{\sigma, \tau},
  \end{aligned}
\end{equation}
where $\phi^{\Lambda}(\boldsymbol{r},\boldsymbol{R})$ and $\phi^{N}(\boldsymbol{r}_{i},\boldsymbol{R}_{i})$ are single particle wave functions of $\Lambda$ particle and nucleons respectively. 
$\chi_{\sigma, \tau}$ is the spin-isospin part of the wave functions and $\boldsymbol{R}$ are generator coordinates, which represent the center of Gaussian functions. $\nu_{N}$ and $\nu_{\Lambda}$ are width parameters for nucleons and $\Lambda$ particles, respectively.  
$\nu_{N}$ is 0.235 $\rm{fm}^{-2}$, which is adopted from Ref.~\cite{suhara_cluster_2012}, and $\nu_{\Lambda} = \nu_{N}(M_{\Lambda}/{M_{N}})^2$, where $M_{\Lambda}$ and $M_{N}$ represent the masses of the $\Lambda$ particle and the nucleon, respectively.
\subsection{C. Single particle wave function of Hyper-Brink with cluster-breaking (CB-Hyper-Brink)}
By introducing the imaginary parts of the Gaussian centers, we construct the single-particle wave function of the CB-Hyper-Brink.
Consequently, the center of the Gaussian packet, $\boldsymbol{\xi}$, becomes a complex number.
\begin{equation}
  \begin{aligned}
  \phi^{\Lambda, N}(\boldsymbol{r},\boldsymbol{\xi})&=
      \left(\frac{2\nu_{\Lambda, N}}{\pi}\right)^{3/4}
      \exp\left\{
          -\nu_{\Lambda, N}(\boldsymbol{r}-\boldsymbol{\xi})^{2}
      \right\}\chi_{\sigma, \tau},
  \end{aligned}
\end{equation}
The imaginary parts of the Gaussian wave packet represent the high-momentum excitation of the particles.

In the present work, we focus on the $_\Lambda^{12}\rm{B}$ hypernucleus.
It is treated as a four-body model, $\alpha+\alpha+t+\Lambda$, microscopically in the Hyper-Brink wave function, where the $\alpha$ clusters are positioned along the $z$-axis symmetrically, and the $t$ cluster and $\Lambda$ particle move in the full coordinate space.
Therefore, the $\alpha$ clusters and the $t$ cluster are broken in different ways, as explained below.
The real part of the generator coordinates, $\text{Re}[\boldsymbol{\xi}_i]$, for all particles is consistent with the Hyper-Brink model.
The imaginary part of the generator coordinates, $\text{Im}[\boldsymbol{\xi}_i]$, which cause the dissolution of each cluster, are fixed along the $x$-axis.
It should be noted that we introduce imaginary components for two neutrons, while the proton has no imaginary part in the broken $t$ clusters.
The formulae for the generator coordinates of the three clusters are as follows:
\begin{equation}
  \begin{aligned}
    \boldsymbol{\xi}_{i}^{\alpha:p_1,n_1} &= R^\alpha\boldsymbol{e}_z+i\Lambda_i\boldsymbol{e}_x\\
    \boldsymbol{\xi}_{i}^{\alpha:p_2,n_2} &= R^\alpha\boldsymbol{e}_z-i\Lambda_i\boldsymbol{e}_x\\
  \end{aligned}
\end{equation}
and
\begin{equation}
  \begin{aligned}
    \boldsymbol{\xi}_{i}^{t:n_1} &= \boldsymbol{R}^t+i\Lambda_i\boldsymbol{e}_x\\
    \boldsymbol{\xi}_{i}^{t:n_2} &= \boldsymbol{R}^t-i\Lambda_i\boldsymbol{e}_x\\
  \end{aligned}
\end{equation}
where $\Lambda$ is a real parameter that characterizes the dissolution of the cluster.
In the present calculation, we adopt the individual $\Lambda$ parameters for the three clusters.
It should be noted that the directions of the imaginary parts in spin-parallel nucleons are identical in the broken $\alpha$ cluster.
The directions of the imaginary parts in spin-antiparallel nucleons are opposite.

Furthermore, we rotate the intrinsic spin orientations of the two broken $\alpha$ clusters about the $y$-axis, while the intrinsic spin orientations of the three valence nucleons remain fixed,
\begin{equation}
  \begin{aligned}
    \chi_{\alpha_1;\sigma}' &= \boldsymbol{\hat{R}}(\alpha=0,\beta=\theta_1,\gamma=0)\chi_{\alpha_1;\sigma},\\
    \chi_{\alpha_2;\sigma}' &= \boldsymbol{\hat{R}}(\alpha=0,\beta=\theta_2,\gamma=0)\chi_{\alpha_2;\sigma},
  \end{aligned}
\end{equation}
where $\chi_{\alpha_1;\sigma}$ ($\chi_{\alpha_2;\sigma}$) represent the spin part of first (second) broken $\alpha$ cluster. $\Omega=(\alpha,\beta,\gamma)$ are the Euler angles. $\boldsymbol{\hat{R}}(\Omega)=e^{-i\alpha\boldsymbol{\hat{J}}_z}e^{-i\beta\boldsymbol{\hat{J}}_y}e^{-i\gamma\boldsymbol{\hat{J}}_z}$ is the rotation operator.

In this process, we can parameterize the intrinsic CB-Hyper-Brink basis as
\begin{equation}\label{eq:parameters}
    \boldsymbol{\eta} = \{R^\alpha, \boldsymbol{R}^t, \Lambda_{1-3}, \theta_1, \theta_2, \boldsymbol{\xi}^\Lambda\},
\end{equation}
The total number of degrees of freedom is $12$ for each basis state.
This is significantly fewer than the degrees of freedom in AMD.

In the present work, we perform the $K$- and parity-projection (K-VAP) for CB-Hyper-Brink basis during the energy variation \cite{myo_variation_2023,Myo2024}.
The projected CB-Hyper-Brink bases are antisymmetrization of single particle wave functions as
\begin{equation}\label{eq:amd-k}
  \Phi^{K;\pm}=\frac{1}{\sqrt{n!}}\hat{P}^K\hat{P}^\pm
    \phi^{\Lambda}(\boldsymbol{r}_\Lambda,\boldsymbol{\xi}_\Lambda)\mathcal{A}\left\{
      \prod_{k=1}^{n} \phi^{N}(\boldsymbol{r}_{k},\boldsymbol{\xi}_{k})
    \right\},
\end{equation}
where $\mathcal{A}$ and $\hat{P}^\pm$ are the antisymmetrization operator and parity projection operator respectively. $n$ is the number of nucleons. $\hat{P}^K$ is the $K$-projection operator, which can be given as
\begin{equation}
    \hat{P}^K=\frac{1}{2 \pi} \int_0^{2 \pi} d \theta e^{-i K \theta} \hat{R}(\theta).
\end{equation}
The operator $\hat{R}(\theta)$ represents the rotation operator about the principal axis in a body-fixed frame.
After applying this projection, the CB-Hyper-Brink basis can be optimized to achieve a definite $K^\pi$ quantum number.
Additionally, the K-VAP method significantly reduces the model space and improves the efficiency of the CB-Hyper-Brink basis as described in our previous work \cite{Tian2025}.

Here we define the total wave function during the energy variation, which is the superposition of the CB-Hyper-Brink basis.
\begin{equation}\label{total_wave_function}
  \begin{aligned}
    |\Psi^{K;\pm}(\boldsymbol{\eta})\rangle=\sum_{i=1}^m C_i\left|\Phi^{K;\pm(i)}(\boldsymbol{\eta}^{(i)})\right\rangle \text {, }  
  \end{aligned}
\end{equation}
where $C_i$ are the superposition coefficients of each CB-Hyper-Brink basis. 
The coefficients $C_i$ and the energy eigenvalue $E$ are determined by solving the Hill-Wheeler equation
\begin{equation}\label{eq:HW}
  \begin{aligned}
    \sum_j\left\langle\Phi^{K;\pm(i)}(\boldsymbol{\eta})|{H}-E| \Phi^{K;\pm(j)}(\boldsymbol{\eta})\right\rangle C_j=0.
  \end{aligned}
\end{equation}
\subsection{D. Control Neural Network(Ctrl. NN)}
We optimize the total CB-Hyper-Brink wave function using the Control Neural Network, as described in Refs \cite{tian_lambda_2024,cheng_evidence_2024}.
Here, we combine the parameters $\boldsymbol{\eta}$ shown in Eq (\ref{eq:parameters}) for each CB-Hyper-Brink basis set (20 bases per set) along with the energy $E$ into a single vector $\boldsymbol{X}$, which serves as the input for the Control Neural Network.
\begin{equation}
  \begin{aligned}
    \boldsymbol{X}=\left[\boldsymbol{\eta}^{(1)},\cdots,\boldsymbol{\eta}^{(i)},\cdots,\boldsymbol{\eta}^{(20)}, E\right]  = \left[\left\{\boldsymbol{\eta}^{(i)}\right\}, E\right]
  \end{aligned}
\end{equation}
where the energy $E$ is determined by solving the Hill-Wheeler equation (\ref{eq:HW}).
The Ctrl.NN consists of an input layer, two hidden layers, and an output layer, along with the variational parameters $\boldsymbol{\eta}$.
In the first step, the variational parameters are iteratively optimized, and the ground-state energy $E^{\prime}$ is obtained by diagonalizing the basis states.
Subsequently, the energy $E^{\prime}$ is combined with the updated variational parameters to construct the new input for the next iteration of the neural network.
\begin{equation}
  \begin{aligned}
    \boldsymbol{X}^{\prime}=\left[\left\{\boldsymbol{\eta}^{(i)\prime}\right\}, E'\right]
  \end{aligned}
\end{equation}
In each epoch, the neural network explores a new set of generator coordinates and updates the input if the corresponding energy decreases. Simultaneously, the neural network parameters are updated at the end of each epoch through
\begin{equation}
  \begin{aligned}
    \frac{\partial\left(H^{\prime}-H\right)}{\partial W}=\frac{\partial\left(H^{\prime}-H\right)}{\partial \boldsymbol{R}^{\prime}} \frac{\partial \boldsymbol{R}^{\prime}}{\partial W}.
  \end{aligned} 
\end{equation}
In this process, neural networks learn just in time the properties of the quantum state. 
The iteration continues until the energy no longer decreases.
In the second step, once the energy curve converges, the variation of the basis states is considered complete.
For the excited states, we constrain the target energy to different specified values for each set of basis states. In this way, the optimization of each group is restricted to a particular energy region, ensuring that the final set of optimized basis states collectively spans a sufficiently wide range of excitation energies. This strategy allows the model space to adequately cover various energy regions and improves the description of excited states with different structures.
In the final stage, Hamiltonian matrix elements are calculated with the updated basis states after applying the total angular momentum projection ($J$-projection), and then the Hamiltonian matrix is diagonalized to obtain the eigenenergies and eigenstates.


\section{III. Results and Discussions}
\begin{figure*}[htbp] 
  \centering 
  \includegraphics[width=1\textwidth]{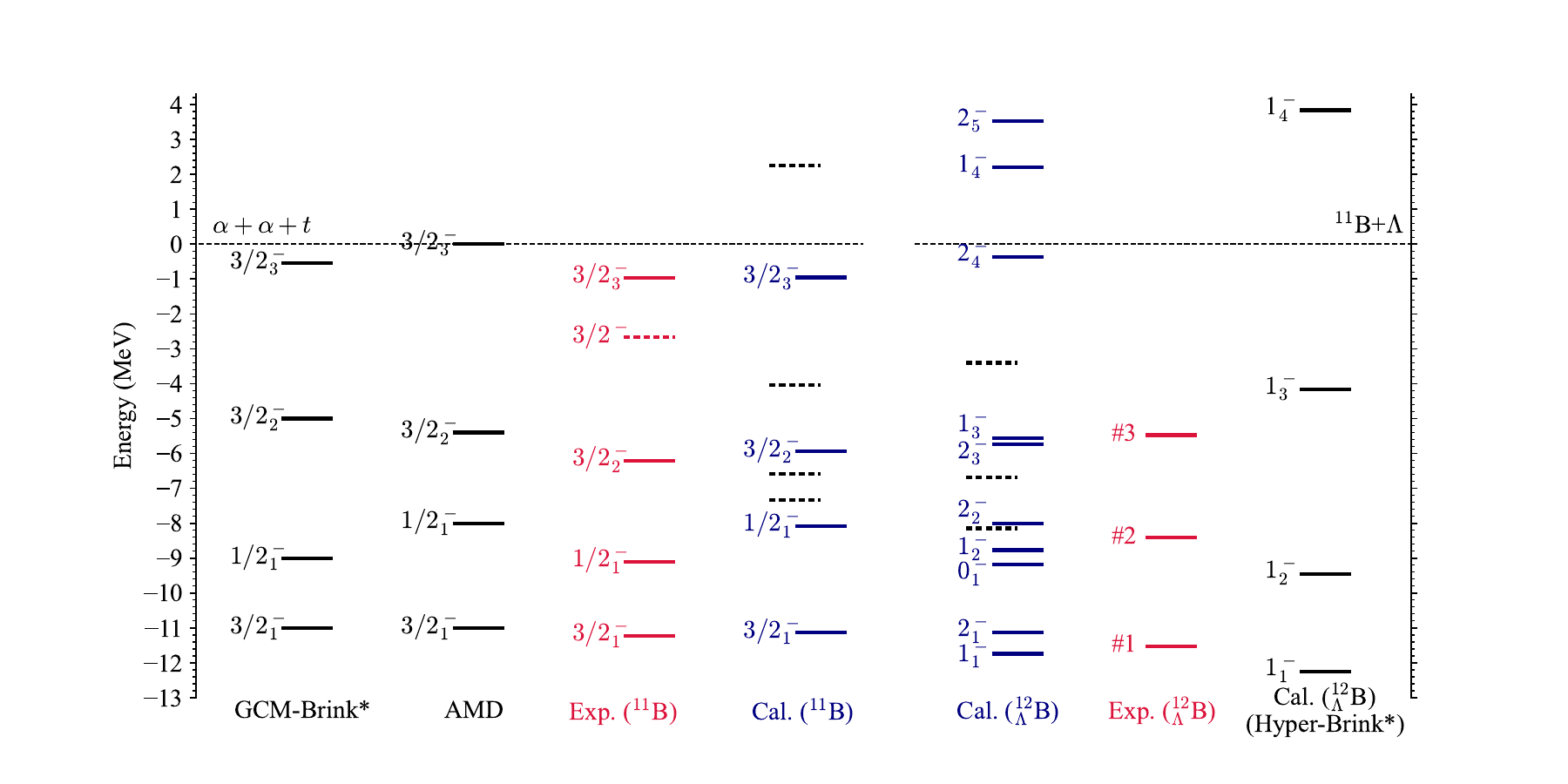} 
  \caption{The spectrum of $^{11}\rm{B}$ and $_\Lambda^{12}\rm{B}$ together with the calculations of GCM-Brink \cite{zhou_2_2018} and AMD \cite{suhara_cluster_2012}. ``Cal. ($^{11}\rm{B}$)" and ``Cal. ($_\Lambda^{12}\rm{B}$)" are  our results obtained with CB-Brink model and CB-Hyper-Brink model, respectively. The dashed lines in present calculations are obtained without cluster breaking (CB) effect. ``Exp. ($^{11}\rm{B}$)" and ``Exp. ($_\Lambda^{12}\rm{B}$)" are the experimental energy levels of $^{11}\rm{B}$ \cite{Kelley2012} and $_\Lambda^{12}\rm{B}$ \cite{tang_experiments_2014}, respectively. An energy level of $^{11}\rm{B}$ with uncertain identification is shown in dashed line. The star indicates that the ``Cluster interaction'' was employed in the GCM-Brink \cite{zhou_2_2018} and Hyper-Brink calculations.} 
  \label{spectrum} 
\end{figure*}
\label{sec:results1}
Before discussing the hypernucleus, we first analyze the results and structures of the $^{11}\rm{B}$ nucleus.
Our focus is on the low-lying states $3/2_1^-$, $1/2_1^-$, $3/2_2^-$, and the Hoyle-analog state $3/2_3^-$.
Fig~\ref{spectrum} presents our calculations alongside experimental data and results from other models where the results with only (Hyper-)Brink bases are also shown with dashed lines.
It should be noted that the GCM-Brink results \cite{zhou_2_2018} employ the ``Cluster interaction", effectively reproducing energy levels from the $\alpha+\alpha+t$ threshold.
By comparing with the results of AMD \cite{suhara_cluster_2012}, it is found that the present model fully incorporates cluster-breaking to describe single-particle motion.
Notably, the Hoyle-analog state $3/2_3^-$ is well reproduced to be about 1 MeV below the $\alpha+\alpha+t$ threshold.
Furthermore, we perform calculations of the energy levels within the cluster model to elucidate the cluster breaking effect.
It is found that this effect is substantial, providing an additional binding energy of about 3 MeV for these states, owing to the manifestation of the spin-orbit interaction in the shell-model-like components.

Table~\ref{radii} presents the root-mean-square (rms) matter radii of these states, comparing various models and experimental data.
The present results are consistent with calculations from other models.
The large radius of the $3/2_3^-$ state indicates a broad spatial distribution, which is a defining characteristic of the Hoyle-analog state.

\begin{table}[htbp]
  \centering
  \renewcommand{\arraystretch}{1.5} 
  \caption{The rms matter radii of negative states of $^{11}\rm{B}$ nucleus  are shown. The results of AMD \cite{suhara_cluster_2012}, GCM-Brink \cite{zhou_2_2018}, AQCM \cite{Itagaki2022} models and experiment \cite{Kelley2012} are also shown. All units are fm.}
  \label{radii}
  \setlength{\tabcolsep}{1.5mm}
  {\begin{tabular}{ccccccccc}
    \hline\hline 
    State & Present &AMD & GCM-Brink & AQCM & Exp. \\
    \hline
    $3/2^-_1$ & 2.30   &2.29 & 2.38  &2.37 &  2.09 $\pm$ 0.12\\
    $1/2^-_1$ & 2.44   & -----  &  2.46   & ----- & 	        \\
    $3/2^-_2$ & 2.50   &2.46 & 2.64   &2.57 & 	\\
    $3/2^-_3$ & 2.78  &2.65 & 2.99  &2.70 & 	\\
    \hline\hline
  \end{tabular}}
\end{table}

To elucidate the cluster-breaking contribution to each state, we compute the expectation values of the one-body spin-orbit operator,
\begin{equation}
  \begin{aligned}
      \mathcal{\hat{O}}^{ls}=\sum_i \hat{\boldsymbol{l}}_i\cdot \hat{\boldsymbol{s}}_i,
  \end{aligned}
\end{equation}
where $\hat{\boldsymbol{l}}_i$ and $\hat{\boldsymbol{s}}_i$ represent the orbital angular momentum and spin of the $i$th nucleon, respectively, and the expectation value is obtained by summing over all nucleons.
In the pure cluster (Hyper-)Brink model, the expectation value is obtained to approach zero.
On the other hand, in the shell model, the expectation value for the $p_{3/2}$ subshell closure configuration should be 3.5, calculated as 7 (the number of nucleons occupying the $p_{3/2}$ orbit) $\times$ 0.5 (the expectation value of $\hat{\boldsymbol{l}}\cdot \hat{\boldsymbol{s}}$ for the $p_{3/2}$ orbit).
It should be noted that the expectation value of $\hat{\boldsymbol{l}}\cdot \hat{\boldsymbol{s}}$ for $p_{1/2}$ orbit is -1. 
Therefore, for states involving $p_{1/2}$ orbit excitation, the cluster-breaking effect should be carefully analyzed based on the expectation values of the one-body spin-orbit operator.
\begin{table}[htbp]
  \centering
  \renewcommand{\arraystretch}{1.5} 
  \caption{The expectation values of the one-body spin-orbit operator for the $^{11}\rm{B}$ nucleus and the $_\Lambda^{12}\rm{B}$ hypernucleus are listed. The explicit configurations of the hypernuclear states are also provided. The ``nucl." and ``prot." denote the summed expectation values for nucleons and protons, respectively.}
  \label{LS}
  \setlength{\tabcolsep}{2mm}
  {\begin{tabular}{ccccccccc}
    \hline\hline 
    $^{11}\rm{B}$ & nucl. &prot. & &$_\Lambda^{12}\rm{B}$ ($^{11}\rm{B}\otimes \Lambda$) & nucl. & prot. \\
    \cline{1-3} \cline{5-7}
    $3/2^-_1$ & 1.92   &0.93  &&$1_1^-$ ($3/2^-_1\otimes s_\Lambda$) &1.74 &  0.87 \\
    $1/2^-_1$ & 0.75   &0.32  &&$1_2^-$ ($1/2^-_1\otimes s_\Lambda$) &0.87 & 0.38  \\
    $3/2^-_2$ & 0.22   &-0.18 &&$1_3^-$ ($3/2^-_2\otimes s_\Lambda$) &0.25 & -0.19 \\
    $3/2^-_3$ & 0.99   &0.54  &&$1_4^-$ ($3/2^-_3\otimes s_\Lambda$) &0.86 & 0.50 \\
    \hline\hline
  \end{tabular}}
\end{table}
Table~\ref{LS} shows the expectation values of the one-body spin-orbit operator for the states of $^{11}\rm{B}$ nucleus and the $_\Lambda^{12}\rm{B}$ hypernucleus, where ``nucl." and ``prot." denote the sums of expectation values for nucleons and protons, respectively.
The results indicate that the degree of cluster breaking in the hypernuclear states of $_\Lambda^{12}\rm{B}$ is overall comparable to that in $^{11}\rm{B}$.
The ground state ($3/2_1^-$) of $^{11}\rm{B}$ can be considered an intermediate state between the pure cluster state and the $p_{3/2}$ subshell closure state, resembling the ground state of $^{12}\rm{C}$ \cite{itagaki_simplified_2005, Suhara2013}.
By evaluating the overlaps between the basis states and the total wave function and identifying the dominant components, we find that, in the $1/2^-_1$ and $3/2^-_2$ states, the valence proton is predominantly spin-down. This configuration promotes its excitation to the $p_{1/2}$ orbit, where the spin-orbit interaction acts repulsively, consistent with Ref.~\cite{Itagaki2022}.
This trend is corroborated by the reduced expectation values of the one-body spin-orbit operator for both nucleons and protons.
Furthermore, the expectation value of the one-body spin-orbit operator for the Hoyle-analog state ($3/2_3^-$) is 0.99, indicating slight cluster-breaking despite its broad spatial distribution.
However, there is a possibility that strong cluster-breaking and pure cluster configurations coexist, forming a state with a modest expectation value of the one-body spin-orbit operator.
Therefore, further clarification of the cluster-shell coexistence and structural characteristics is necessary.

\begin{figure}[htbp] 
  \centering 
  \includegraphics[width=0.45\textwidth]{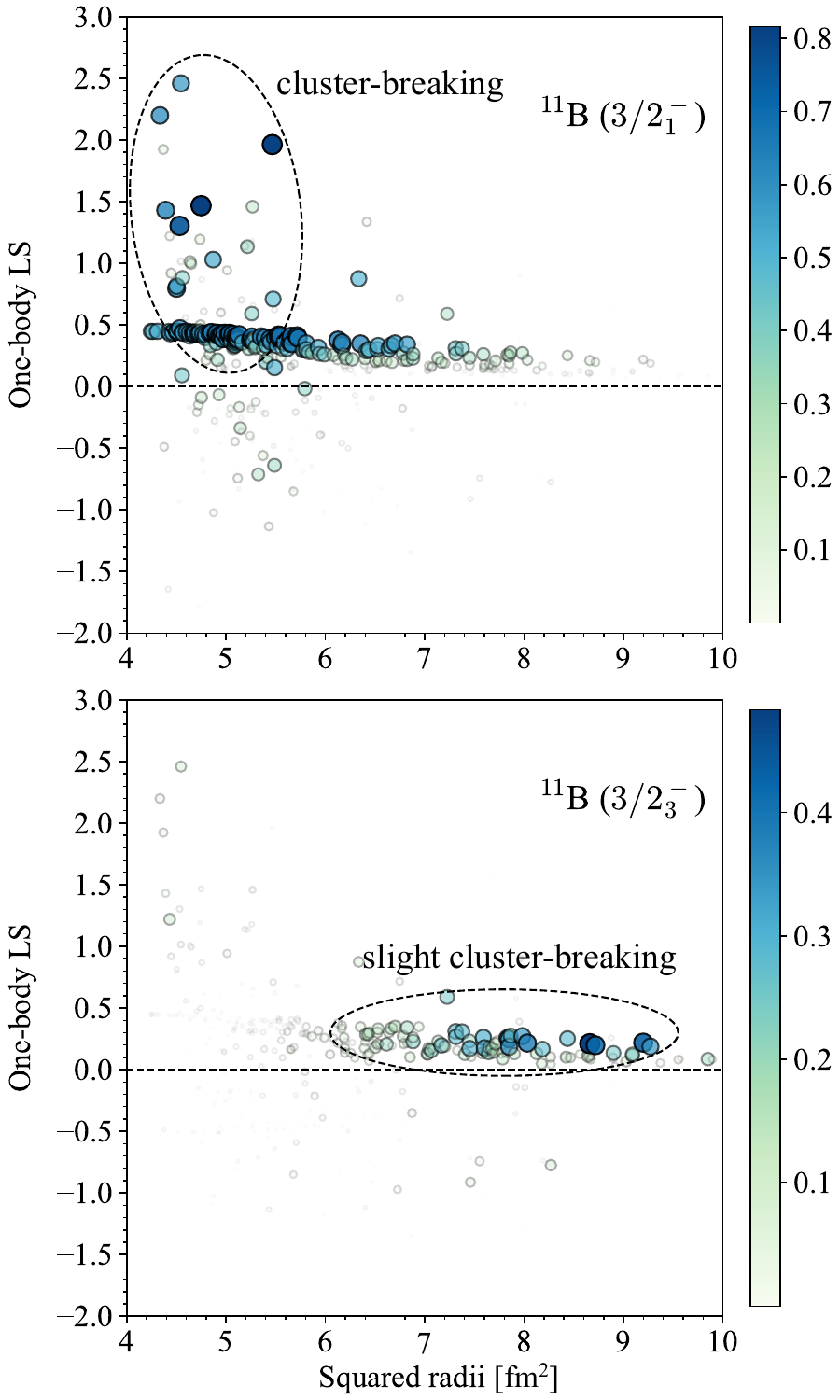} 
  \caption{The scatter plots showing the overlap between each basis state and the $3/2_1^-$ ($3/2_3^-$) state of $^{11}\rm{B}$ are presented in the upper (lower) panel. The color, size, and transparency collectively represent the value of the overlap. The horizontal and vertical axes represent the expectation value of the one-body spin-orbit operator $\mathcal{\hat{O}}^{ls}$ and the squared radii $\hat{R^2}$ for each basis state in units of fm$^2$, respectively.} 
  \label{R-LS} 
\end{figure}

To achieve this, we investigate the relationship among the expectation value of the one-body spin-orbit operator, radii, and the main bases of states, as shown in Fig~\ref{R-LS}.
During the energy variation process for multi-Slater determinants by Ctrl.NN, similarly to the multi-cool method \cite{myo_variation_2023,Myo2024}, the basis states are generated to characterize the states of the $^{11}\rm{B}$ nucleus and the $_\Lambda^{12}\rm{B}$ hypernucleus.
We calculate the squared overlap between each basis state and the total wave function, which is defined in Refs~\cite{tian_lambda_2024, Tian2025}.
The scatter plots of overlap for each basis state describing $^{11}\rm{B}$ are presented in Fig~\ref{R-LS}, with the color, size, and transparency representing the overlap values.
As shown in Fig.~\ref{R-LS}, the expectation values of the one-body spin-orbit operator for most basis states describing the $3/2^-_1$ state of $^{11}\mathrm{B}$ deviate significantly from the pure cluster-model value of zero. This indicates strong cluster breaking in the ground state ($3/2^-_1$) of $^{11}\mathrm{B}$.
Notably, the Hoyle-analog state ($3/2_3^-$) exhibits a broad spatial distribution with slight cluster-breaking.

Next, we focus on the results of these states after introducing a $\Lambda$ particle.
Here, we select the hypernuclear states in which the $\Lambda$ particle dominantly occupies the $s$ state, following the same method described in the Appendix A of Ref.~\cite{Tian2025}.
The energy levels are shown in Fig.~\ref{spectrum}.
Additionally, we present the results of the Hyper-Brink cluster model for $1^-$ states by using the ``Cluster interaction."
Significant differences are observed between the two models, even though the energy levels of the core nuclei are well reproduced.
This indicates that cluster-breaking plays a crucial role in describing the hypernuclear wave function, particularly for low-lying shell-like states.
Due to the $2^-$ states of $_\Lambda^{12}\mathrm{B}$ include the core nucleus excitation of $^{11}\mathrm{B}(5/2^-)$, we concentrate on the discussion of $1^-$ states.
The level spacing between the ground state (shell-like) and the Hoyle-analog state increases upon the addition of a $\Lambda$ hyperon. In our calculation, the spacing between the $1^-_1$ and $1^-_4$ states of $_\Lambda^{12}\mathrm{B}$ exceeds that between the $3/2^-_1$ and $3/2^-_3$ states of $^{11}\mathrm{B}$ by 3.8 MeV.
This result closely resembles the enhanced energy level of the Hoyle state, $^{12}\rm{C}$ ($0_2^+$), as reported in Ref~\cite{Hiyama2000}.
To further investigate the importance of cluster-breaking, we calculate the electric quadrupole transition strength $B(E2)$, defined by the corresponding transition operator and strength as
\begin{equation}
  M_\mu^{E2} = e \sum_{i=1}^A (\boldsymbol{r}_i-\boldsymbol{r}_{cm})^2Y_{2\mu}(\Omega_{\boldsymbol{r}_i-\boldsymbol{r}_{cm}})\frac{1+\tau_{iz}}{2},
\end{equation} 
\begin{equation}
  B(E2;J_i\rightarrow J_f) = \sum_{M_f \mu} |\braket{J_f M_f |M_\mu^{E2}|J_i M_i}|^2,
\end{equation} 
where the $\tau_{iz}$ represents the isospin projection of the $i$th nucleon.
To compare the $B(E2)$ strengths of $_\Lambda^{12}\rm{B}$ and $^{11}\rm{B}$, we apply a correction to $B(E2)$, assuming that the $\Lambda$ particle occupies the $s$ orbit in the low-lying hypernuclear states.
This treatment of $B(E2)$ is consistent with previous studies in Refs~\cite{motoba_progr_1983,Hiyama1999,isaka_impurity_2015,Isaka2012}.
The $B(E2)$ results are presented in Table~\ref{be2}, alongside other model calculations.
The current $B(E2)$ results are consistent with those from the AMD calculation.
While the energy levels obtained from the GCM-Brink calculation with "Cluster interaction" align well with the current results, the $B(E2)$ value shows a notable discrepancy.
By comparing the $3/2^-_3 \rightarrow 3/2^-_1$ transition in $^{11}\mathrm{B}$ with the $1^-_4 \rightarrow 1^-_1$ transition in $_{\Lambda}^{ 12}\mathrm{B}$, the present calculation shows that the introduction of a $\Lambda$ hyperon enhances the $B(E2)$ value by approximately 36$\%$, whereas it remains nearly unchanged in the Hyper-Brink cluster model.
The different change in $B(E2)$ between these two models serves as a probe to assess the importance of cluster-breaking in describing hypernuclear states, based on precise experimental data in the future.

\begin{table}[htbp]
  \centering
  \renewcommand{\arraystretch}{1.5} 
  \caption{The $B(E2)$ values in the $^{11}\rm{B}$ nucleus and the corresponding transition in $_\Lambda^{12}\rm{B}$ are presented. The values obtained from AMD \cite{Kawabata2007} and GCM-Brink \cite{zhou_2_2018} models are also provided. The unit is $e^2$fm$^4$. The star indicates that the ``Cluster interaction'' was employed in the energy level calculations. The $cB(E2)$ represents the corrected $B(E2)$ values in $_\Lambda^{12}\rm{B}$, with the definition of corrected $B(E2)$ values being consistent with Ref.~\cite{motoba_progr_1983,Hiyama1999,isaka_impurity_2015,Isaka2012}.The column ``Changes (\%)'' denotes the relative change of $cB(E2)$ in $_\Lambda^{12}\rm{B}$ with respect to the corresponding $B(E2)$ in $^{11}\rm{B}$ for each paired transition.}
  \label{be2}
  \setlength{\tabcolsep}{2.3mm}
  {\begin{tabular}{ccccccc}
    \hline\hline 
     &$^{11}\rm{B}$ & $B(E2)$ &\\
    \hline 
    Present&$3/2^-_3\rightarrow3/2^-_1$ & 0.74&\\
          &$3/2^-_2\rightarrow3/2^-_1$ & 0.01&\\
          &$3/2^-_3\rightarrow3/2^-_2$ &0.06&\\
    AMD&$3/2^-_3\rightarrow3/2^-_1$ & 0.84&\\
      &$3/2^-_2\rightarrow3/2^-_1$ & 0.02&\\
    GCM-Brink*&$3/2^-_3\rightarrow3/2^-_1$ & 1.5&\\
      &$3/2^-_2\rightarrow3/2^-_1$ & 0.07&\\
    \hline 
     &$_\Lambda^{12}\rm{B}$ & $cB(E2)$ &Changes ($\%$)\\
    \hline 
    Present&   $1_4^-\rightarrow1_1^-$ & 1.01&  +36\\
          &   $1_3^-\rightarrow1_1^-$ & 0.020&  +107 \\
          &   $1_4^-\rightarrow1_3^-$ & 0.016&   -74\\
    Hyper-Brink*&$1_4^-\rightarrow1_1^-$ & 1.49& -7\\
            &   $1_3^-\rightarrow1_1^-$ & 0.003 &-71\\
    \hline\hline
  \end{tabular}}
\end{table}
\begin{figure}[htbp] 
  \centering 
  \includegraphics[width=0.45\textwidth]{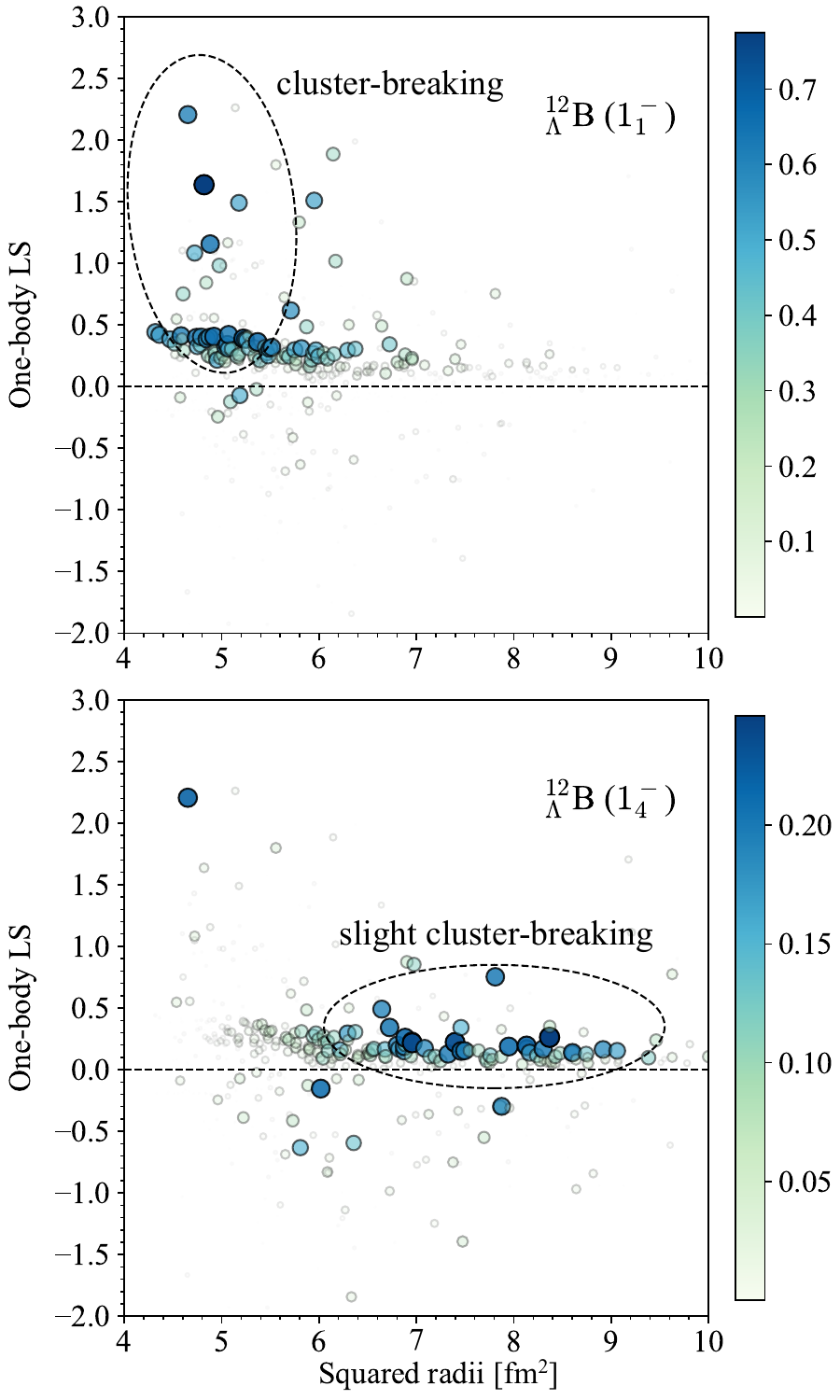} 
  \caption{The scatter plots showing the overlap between each basis state and the $1_1^-$ and $1_4^-$ states of $_\Lambda^{12}\rm{B}$ are presented, similarly to Fig.~\ref{R-LS}.} 
  \label{R-LS-hyper} 
\end{figure}

In a similar way to the previous analysis, we now discuss cluster-breaking based on the expectation values of the one-body spin-orbit operator.
By comparing the result of $_\Lambda^{12}\rm{B}$ with that of $^{11}\mathrm{B}$ in Table~\ref{LS}, slight reductions of the expectation values of the one-body spin-orbit operator are observed in the $1_1^-$ and $1_4^-$ states, while slight enhancements occur in the $1_2^-$ and $1_3^-$ states.
Here, we focus on the ground state and the Hoyle-analog state with the introduction of the $\Lambda$ particle and present the relationship between the expectation value of the one-body spin-orbit operator, radii, and the main bases of the states in Fig.~\ref{R-LS-hyper}.
Notably, by comparing with Fig.~\ref{R-LS}, the radius distributions of the $^{12}_{\Lambda}\mathrm{B}$ basis functions are shifted toward smaller values relative to those of $^{11}\mathrm{B}$, in agreement with the shrinkage effect induced by the $\Lambda$ particle.
Furthermore, the differences in the expectation values of the one-body spin-orbit operator in single basis states are negligible when comparing the states of $^{11}\rm{B}$.
However, taking into account the values calculated with total wave functions, the slight reductions shown in Table~\ref{LS} suggest a minor enhancement in clustering.
\begin{table}[htbp]
  \centering
  \renewcommand{\arraystretch}{1.5} 
  \caption{The rms matter radii of $3/2_1^-$, $3/2_3^-$ states of $^{11}\rm{B}$ nucleus and corresponding hypernuclear states of $^{12}_{\Lambda}\mathrm{B}$ are shown. The core nuclear rms matter radii of hypernuclear states are enclosed in parentheses.}
  \label{radii-hyper}
  \setlength{\tabcolsep}{3.2mm}
  {\begin{tabular}{ccccc}
    \hline\hline 
    $^{11}\rm{B}$ & radii & &$_\Lambda^{12}\rm{B}$ ($^{11}\rm{B}\otimes \Lambda$) & radii \\
    \cline{1-2} \cline{4-5}
    $3/2^-_1$ & 2.30   &&$1_1^-$ ($3/2^-_1\otimes s_\Lambda$) &2.26 (2.28)  \\
    $3/2^-_3$ & 2.78   &&$1_4^-$ ($3/2^-_3\otimes s_\Lambda$) &2.52 (2.54) \\
    \hline\hline
  \end{tabular}}
\end{table}

Here, we discuss the reasons for the slightly enhanced clustering in the ground state $1_1^-$ and the Hoyle-analog state $1_4^-$.
The root-mean-square (rms) matter radii for these two states are listed in Table~\ref{radii-hyper}.
Furthermore, we calculate the squared overlap between the (hyper-)nuclear total wave functions, ${\Psi^{J \pi}_\text{total} }$, and the $J^\pi$-projected (Hyper-)Brink basis functions, ${\hat{P}^{J \pi}_{MK} \Psi_B}$, to analyze the spatial distribution and clustering,
\begin{equation}\label{eq:overlap-brink}
  U(R_{\alpha\alpha}, \boldsymbol{R}^t, \boldsymbol{R}^\Lambda) = |\braket{\hat{P}^{J\pi}_{MK}\Psi_{B}(R_{\alpha\alpha}, \boldsymbol{R}^t, \boldsymbol{R}^\Lambda)|\Psi^{J\pi}_\text{total}}|^2.
\end{equation} 
In the $3/2^-_1$ and $3/2^-_3$ states of $^{11}\rm{B}$ and the $1^-_1$ and $1^-_4$ states of $_\Lambda^{12}\rm{B}$, the quantum numbers $K = 3/2$ and $K = 1$, respectively, are chosen.
The distance between the two $\alpha$ clusters, $R_{\alpha\alpha}$, and the generator coordinate of $\Lambda$, $\boldsymbol{R}^\Lambda$, are fixed to optimal values that maximize the squared overlap.
The generator coordinate of the triton cluster, $\boldsymbol{R}^t$, is fixed on the $y = 0$ cross section.
\begin{figure}[htbp] 
  \centering 
  \includegraphics[width=0.48\textwidth]{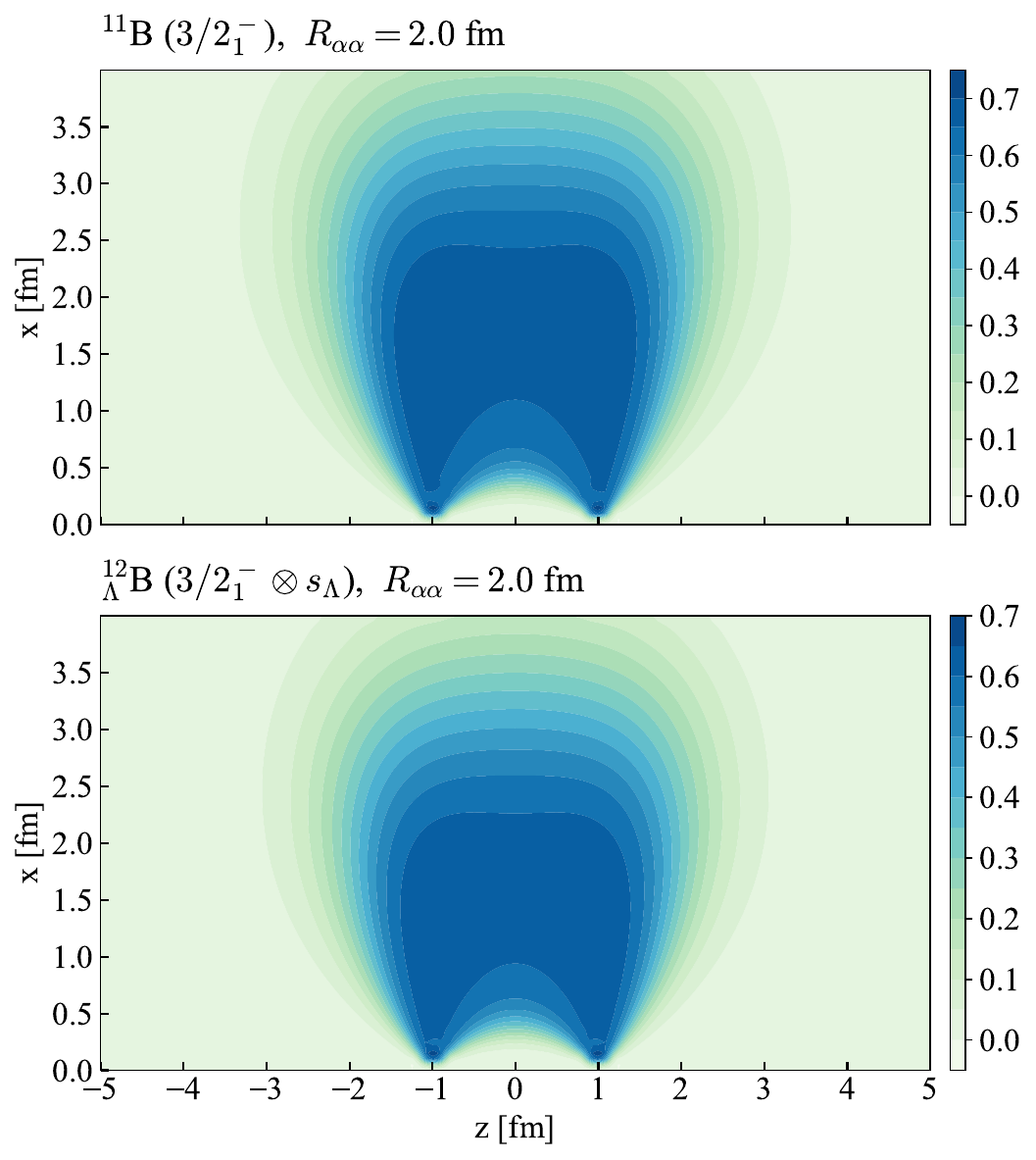} 
  \caption{The contour plots of the squared overlap $U(R_{\alpha\alpha}, \boldsymbol{R}^t, \boldsymbol{R}^\Lambda)$, as defined in Eq.~(\ref{eq:overlap-brink}), for $^{11}\rm{B}$($3/2^-_1$) and $_\Lambda^{12}\rm{B}$($1^-_1$) are presented as functions of $\boldsymbol{R}^t$. The horizontal and vertical axes represent the $z$ and $x$ components of $\boldsymbol{R}^t$, respectively. The distance between the two $\alpha$ clusters is fixed at an optimal value to maximize the squared overlap.} 
  \label{overlap-ground} 
\end{figure}

For the ground states, $R_{\alpha\alpha}$ are fixed at 2 fm and $\boldsymbol{R}^\Lambda$ is positioned at the center of the coordinate space.
The contour plots of the squared overlap $U(R_{\alpha\alpha}, \boldsymbol{R}^t, \boldsymbol{R}^\Lambda)$ as defined in Eq.~(\ref{eq:overlap-brink}) for the ground states of $^{11}\rm{B}$ and $_\Lambda^{12}\rm{B}$ are presented in Fig.~\ref{overlap-ground}.
The introduction of the $\Lambda$ particle leads to a very slight shrinkage of the nuclear core size, due to the short-range repulsion between the $\Lambda$ particle and nucleons and the incompressibility of the nuclear core.
The interplay between short-range repulsion and intermediate attraction in the $\Lambda$-nucleon interaction leads to a slight aggregation of valence nucleons.

\begin{figure}[htbp] 
  \centering 
  \includegraphics[width=0.48\textwidth]{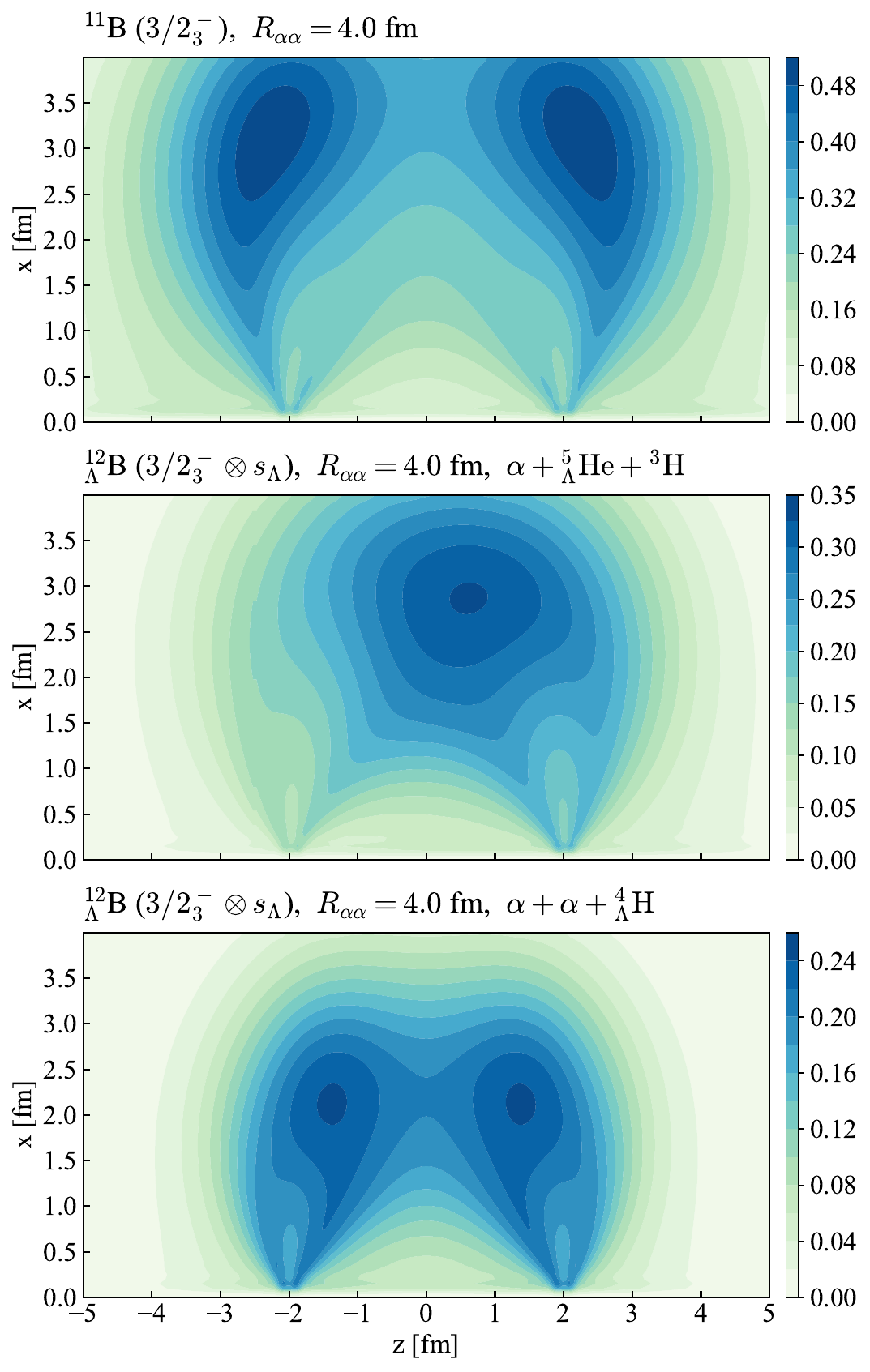} 
  \caption{The contour plots of the squared overlap $U(R_{\alpha\alpha}, \boldsymbol{R}^t, \boldsymbol{R}^\Lambda)$ are presented for $^{11}\rm{B}$($3/2^-_3$) and $_\Lambda^{12}\rm{B}$($1^-_4$) as functions of $\boldsymbol{R}^t$, similarly to Fig~\ref{overlap-ground}. For $_\Lambda^{12}\rm{B}$($1^-_4$), the two configurations $\alpha+ {_\Lambda^{5}\rm{He}}+ {^{3}\rm{H}}$ and $\alpha+ \alpha+ {_\Lambda^{4}\rm{H}}$ are considered.} 
  \label{overlap-hoyle} 
\end{figure}
For the Hoyle-analog state, $R_{\alpha\alpha}$ is fixed at 4 fm. 
The position of $\boldsymbol{R}^\Lambda$ is determined using two configurations: $\alpha+ {_\Lambda^{5}\rm{He}}+ {^{3}\rm{H}}$ and $\alpha+ \alpha+ {_\Lambda^{4}\rm{H}}$, respectively, due to the attractive correlation between the $\Lambda$ particle and clusters.
The contour plots of the squared overlap $U(R_{\alpha\alpha}, \boldsymbol{R}^t, \boldsymbol{R}^\Lambda)$ as defined in Eq.~(\ref{eq:overlap-brink}) for the Hoyle-analog states are presented in Fig.~\ref{overlap-hoyle}.
Notably, a significant shrinkage of valence nucleons is observed in both configurations.
Furthermore, the maximum overlap values are comparable in the two configurations. This indicates that the cluster reconfiguration effect, which is induced by the $\Lambda N$ interaction and characterized by the coexistence of $\Lambda$-$\alpha$ and $\Lambda$-triton correlations in the Hoyle-analog state of $_\Lambda^{12}\rm{B}$, subtly enhances the stability of clusterized configurations through the interplay of short-range repulsion and intermediate-range attraction.

In addition, to verify the model independence of this conclusion, we have performed the same analysis with two other $\Lambda N$ interactions, YNG-ESC16$^+$ \cite{Nagels_ESC16_2019} and YNG-JA \cite{yamamoto_hyperon-nucleon_1994}. The resulting spatial distributions and dominant configurations remain essentially unchanged, and the reconfiguration effect persists in both cases. These supplementary calculations confirm that the cluster reconfiguration mechanism is robust and does not depend sensitively on the specific choice of the $\Lambda N$ interaction.
\section{IV. conclusion}\label{conclusion1}
We investigate the negative-parity states and cluster-breaking effects of the hypernucleus $^{12}_\Lambda$B, with a focus on its ground state and Hoyle-like excited state, using the Hyper-Brink model with cluster-breaking (CB-Hyper-Brink) optimized by Control Neural Network (Ctrl.NN) method.
Benefiting from the inclusion of cluster-breaking effects, our model successfully reproduces the observed low-lying energy levels, thereby demonstrating a reliable prediction of the Hoyle-analog state $1_4^-$ in $^{12}_\Lambda$B. 
Analysis of the one-body spin-orbit operator expectation values reveals strong cluster-breaking effects for both the ground and Hoyle-analog states of $^{12}_\Lambda$B.
Moreover, the calculation of the overlap between the hypernuclear total wave functions and the $J^\pi$-projected (Hyper-)Brink basis functions reveals a clear shrinkage effect in the Hoyle-analog state of $^{12}_\Lambda$B. This shrinkage and the enhanced stability of cluster structures are attributed to the cluster reconfiguration effect, which is induced by the $\Lambda N$ interaction and characterized by the coexistence of $\Lambda$-$\alpha$ and $\Lambda$-triton correlations.
Furthermore, the calculation of the electric quadrupole transition strength, $B(E2)$, for the negative-parity states of $^{12}_\Lambda$B demonstrates that its variation between the ground and Hoyle-analog states serves as a crucial probe for assessing the importance of cluster-breaking effects in describing hypernuclear dynamics, pending future experimental validation.

\section{acknowledgement}
This work was supported by National Natural Science 
Foundation of China [Grant Nos. 12575125, 12205105, 12305123], 
by the National Key R\&D Program of China (Contract 
No.2023YFA1606503), by the China Scholarship Council (CSC) under Grant No. 202406830152, by the JSPS KAKENHI Grant No.JP22K03643 and JP25H01268, by JST ERATO Grant No.JPMJER2304. 
This work is partially supported by High Performance Computing Platform of Nanjing University of Aeronautics and Astronautics.

\bibliographystyle{apsrev4-1}
\bibliography{paper-B12-bib}
\end{document}